\documentclass[a4paper,twoside]{article}

\usepackage{epsfig}
\usepackage{subcaption}
\usepackage{calc}
\usepackage{amssymb}
\usepackage{amstext}
\usepackage{amsmath}
\usepackage{amsthm}
\usepackage{multicol}
\usepackage{pslatex}
\usepackage{apalike}
\usepackage{tabularx}
\usepackage[bottom]{footmisc}
\usepackage{tikz}
\usetikzlibrary{quantikz}
\usepackage{adjustbox}
\usepackage{hyperref}

\usepackage{todonotes}

\usepackage{SCITEPRESS}     

\begin{document}

\title{Improving Convergence for Quantum Variational Classifiers using Weight Re-Mapping}

\author{\authorname{
Michael Kölle\sup{1}, 
Alessandro Giovagnoli\sup{1}, 
Jonas Stein\sup{1}, 
Maximilian Balthasar Mansky\sup{1}, 
Julian Hager\sup{1}, 
Claudia Linnhoff-Popien\sup{1} }
\affiliation{\sup{1}Institute of Informatics, LMU Munich, Oettingenstraße 67, Munich, Germany}
\email{\{michael.koelle, jonas.stein, maximilian-balthasar.mansky, julian.hager\}@ifi.lmu.de, linnhoff@lmu.de}
}


\abstract{
In recent years, quantum machine learning has seen a substantial increase in the use of variational quantum circuits (VQCs). VQCs are inspired by artificial neural networks, which achieve extraordinary performance in a wide range of AI tasks as massively parameterized function approximators. VQCs have already demonstrated promising results, for example, in generalization and the requirement for fewer parameters to train, by utilizing the more robust algorithmic toolbox available in quantum computing.
A VQCs' trainable parameters or weights are usually used as angles in rotational gates and current gradient-based training methods do not account for that. 
We introduce weight re-mapping for VQCs, to unambiguously map the weights to an interval of length $2\pi$, drawing inspiration from traditional ML, where data rescaling, or normalization techniques have demonstrated tremendous benefits in many circumstances. We employ a set of five functions and evaluate them on the Iris and Wine datasets using variational classifiers as an example.
Our experiments show that weight re-mapping can improve convergence in all tested settings. Additionally, we were able to demonstrate that weight re-mapping increased test accuracy for the Wine dataset by $10\%$ over using unmodified weights.}

\onecolumn \maketitle \normalsize \setcounter{footnote}{0} \vfill

\section{\uppercase{Introduction}}
\label{sec:introduction}
Machine learning (ML) tasks are ubiquitous in a vast number of domains, including central challenges of humanity, such as drug discovery or climate change forecasting. In particular, ML techniques allow to tackle problems, for which computational solutions were unimaginable before its rise. While many tasks can be solved efficiently using machine learning techniques, fundamental limitations are apparent in others, e.g., the curse of dimensionality \cite{bellman1966dynamic}.
Inspired by the success of classical ML and a promising prospect to circumvent some of its intrinsic limitations, quantum machine learning (QML) has become a central field of research in the area of quantum computing. Based on the laws of quantum mechanics rather than classical mechanics, a richer algorithmic tool set can be used to solve some computational problems faster on a quantum computer than classically possible. In some special cases, this new technology allows for exponential speedups for specific problems, such as classification and regression, i.e., by employing a quantum algorithm to solve systems of linear equations (SLEs), whose runtime is logarithmic in the dimensions of the SLEs for sparse matrices \cite{Harrow_2009}.
Aside from the application of such provably efficient basic linear algebra subroutines, a central pillar of quantum computing, a new field in QML has recently been proposed: Variational Quantum Computing (VQC). Using elementary constructs of quantum computing, a universal function approximator can be constructed, much in the same way as in artificial neural networks. In analogy to logic gates, quantum gates are the fundamental algorithmic building blocks. The mathematical operations represented by quantum computing building blocks are different from classical elements, as they describe rotations and reflections, or concatenations of these in a vector space. Rather than working in a general vector space, quantum computing acts on a Hilbert space with exponentially more dimensions available with increasing number of parameters. In essence, every variational quantum circuit can be decomposed into a set of parametrized single qubit rotations and unparmeterized reflections \cite{nielsen_chuang_2010}.

While variational quantum computing has benefits such as fewer training parameters to optimize \cite{Du_2020}, many open problems remain in its practical application. The structure of a quantum circuit is its own problem, where many different concatenations of quantum gates are possible and lead to different training results. At the same time, the Hilbert space is a closed space with possible values from $[0,2\pi]$. It is still unclear how to best work with the classically unusual domain restriction of the parameters, set by the rotation operations.
This problem is of great relevance, as rotations, being periodic functions, are not injective, and thus lead to ambiguous parameter assignments and thus worse training performance.

\begin{figure}[t]
    \centering
    \includegraphics[width=\linewidth]{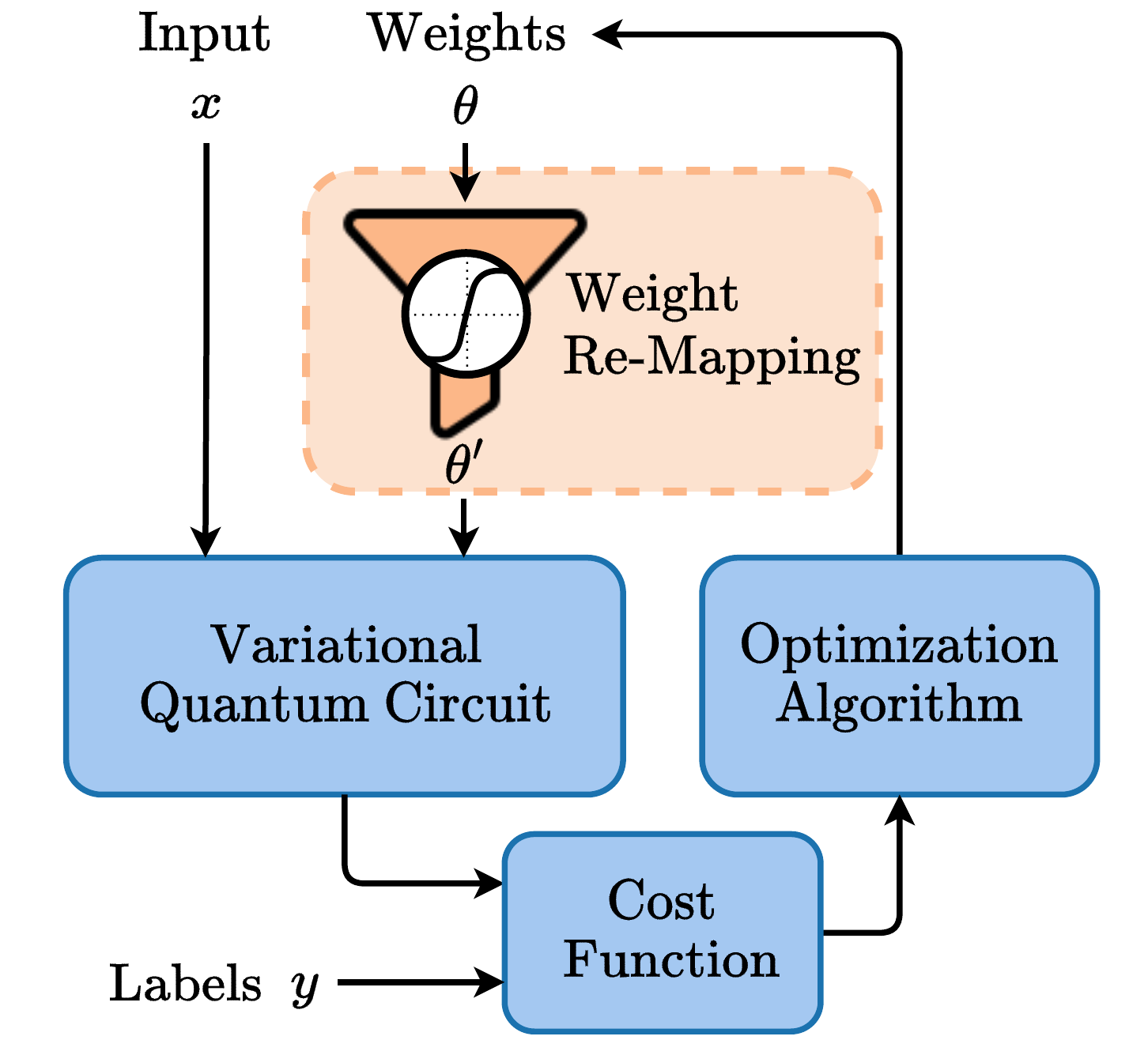}
    \caption{Overview of the Variational Quantum Circuit Training Process with Weight Constraints}
    \label{fig:structure}
\end{figure}

Drawing inspiration from classical ML, where data rescaling, or normalization techniques have shown immense improvements in many cases \cite{Singh2019}, we propose the introduction of re-mapping training parameters in variational quantum circuits as described in Figure \ref{fig:structure}. Concretely, we employ a set of well known fixed functions to unambiguously map the weights to an interval of length $2\pi$ and test their performance. For evaluation purposes, we use a classification problem and a circuit architecture suitable for the employed variational classifier.
Our experimental data shows, that the proposed weight re-mapping leads to faster convergence in all tested settings compared to runs with unconstrained weights. In some cases, the overall test accuracy can also be improved.
In this work, we first describe the basics of variational quantum circuits and related work. We then explain the idea behind our approach and how we setup up our experiments. Finally, we present and discuss our results and end with a summary, limitations and future work. All experiments and a PyTorch implementation of the used weight re-mapping functions can be found here \footnote{\url{https://github.com/michaelkoelle/qw-map}}.

\section{Variational Quantum Circuits}
The most prominent function approximator used in classical machine learning is the artificial neural network: a combination of parameterized linear transformations and typically non-linear activation functions, applied to neurons. The weights and biases used to parameterize the linear transformations can be updated using gradient based techniques like backpropagation, optimizing the approximation quality. According to Cybenko's universal approximation theorem, this model allows the approximation of arbitrary functions with arbitrary precision.\cite{Cybenko1989}.

In a quantum circuit, information is stored in the state of a qubit register $\ket{\psi_i}$, i.e., normalized vectors living in a Hilbert space $\mathcal{H}$. In quantum mechanics, a function mapping the initial state $\ket{\psi_i}$ onto the final state $\ket{\psi_f}$ is expressed by a unitary operator $U$ that maps the inputs onto the outputs as in $\ket{\psi_f} = U \ket{\psi_i}$. In contrast to classical outputs, quantum outputs can only by obtained via so called measurements, which yield an eigenstate corresponding with an expected value of $\bra{\psi_f}O\ket{\psi_f}$, where $O$ is typically chosen to be the spin Hamiltonian in the $z$-axis.

In order to build a quantum function approximator in form of a VQC, one typically decomposes the arbitrary unitary operator $U$ into a set of quantum gates. Analogously to Cybenko's theorem, in the quantum case it can be proved that any unitary operator acting on multiple qubits can always be expressed by the combination of controlled-not (CNOT) and rotational (ROT) gates, which represent reflections or rotations of the vector into the Hilbert space respectively \cite{nielsen_chuang_2010}. While CNOTs are parameter-free gates, each rotation is characterized by the three angles around the axes of the Bloch sphere. These rotation parameters are the weights of the quantum variational circuit. We can thus say that the final state actually depends on the weights $\theta$ of the circuit, and rewrite the output final value as 
\begin{equation}
    \bra{\psi_f(\theta)}O\ket{\psi_f(\theta)}
\end{equation}
Starting from this theoretical basis, a function approximator can be obtained once a suitable circuit structure, also called ansatz, has been chosen. Once this is done and an objective function has been chosen, the rotation weights can be trained in a quantum-classical pipeline, as shown in Figure \ref{fig:structure}, completely analogously to what is done with a neural network.

Similar to a classical neural network, where the gradient is calculated using backpropagation, we can differentiate the circuit with respect to the parameters $\theta$ in a similar way using the parameter shift rule \cite{schuld_evaluating_2019}. It has to be noted however, that this approach has linear time complexity in the number of parameters and every gradient calculation has to be executed on a quantum computer. This is significantly worse than the constant time complexity regarding the number of parameters, needed in backpropagation.
\section{\uppercase{Related Work}}

The task of embedding data from $\mathbb{R}^n$ to $SU(2^n)$ has received limited attention so far, with approaches focusing on global embeddings  \cite{lloyd_quantum_2020} or local mappings. The former approach adds another classical calculation overhead to the whole VQC construction. It is also data specific and needs to be retrained for each dataset. 

For local embeddings there are arguments for rescaling the data classically to a fixed interval as $[\min(\text{data}), \max(\text{data})]\to [0,1]$ for each data dimension and then mapping that to rotations on a single qubit axis \cite{stoudenmire_supervised_2016}, multi-qubit embedding \cite{mitarai_quantum_2018} and random linear maps \cite{wilson_quantum_2019,benedetti_parameterized_2019}.
\section{\uppercase{Our Approach}}
In this section, we first explain the idea and the process behind weight re-mapping for variational quantum circuits. Then we elaborate on the chosen example implementation of a variational quantum classifier architecture and the datasets on which we evaluate the approach. Note that, the approach is not limited to supervised learning tasks. Because weight re-mapping can be applied to VQCs that act as function approximators, it can be easily applied to other machine learning tasks such as unsupervised learning, reinforcement learning, and so on. Finally, we detail our training procedure and pre-tests that influenced our architecture decisions.

\subsection{Weight Re-Mapping}
\label{sec:weight-constraints}
\begin{figure*}
     \centering
     \begin{subfigure}[t]{0.16\textwidth}
         \centering
         \includegraphics[width=\textwidth]{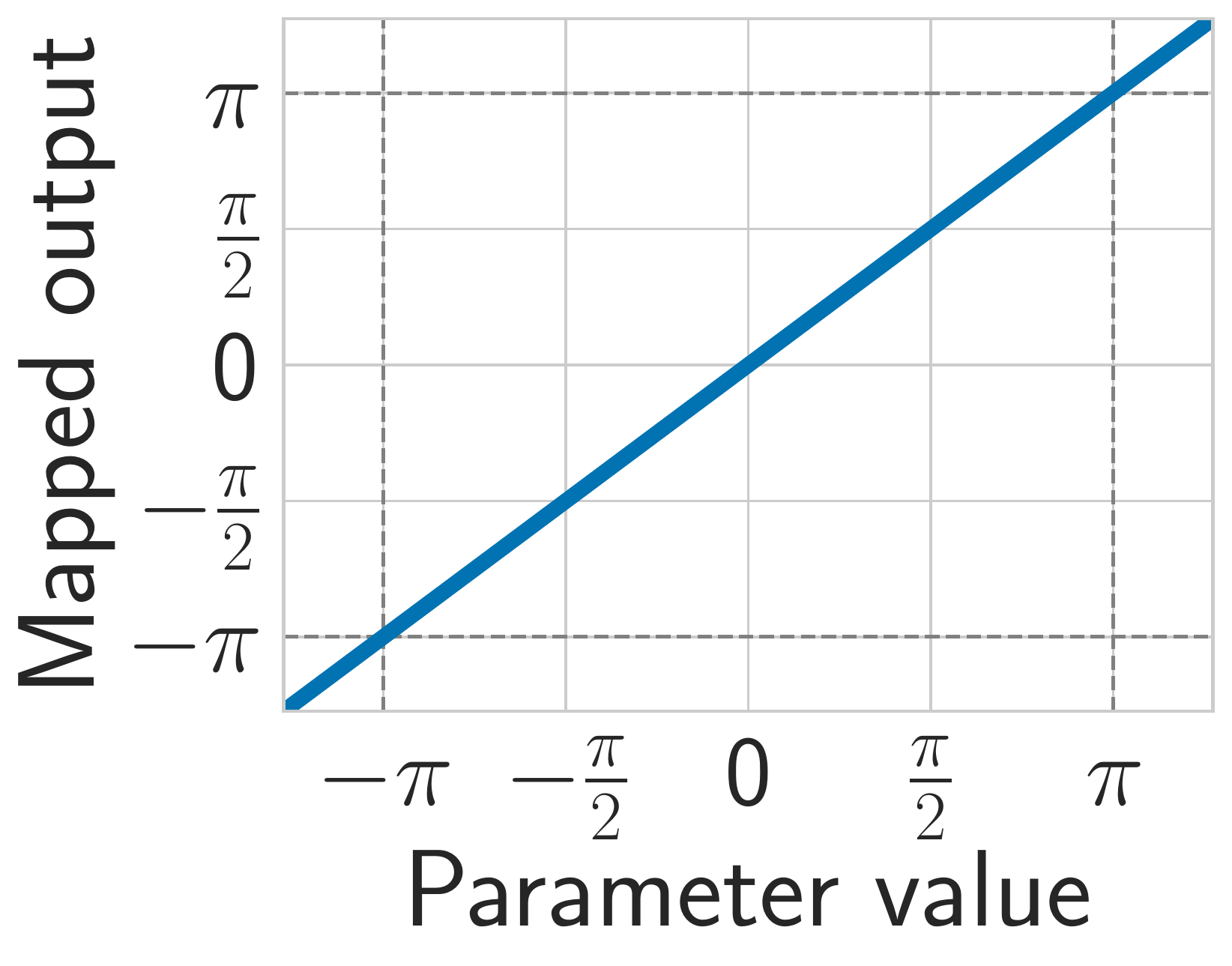}
         \caption{no Re-Mapping}
         \label{fig:func-wo-constraint}
     \end{subfigure}
     \begin{subfigure}[t]{0.16\textwidth}
         \centering
         \includegraphics[width=\textwidth]{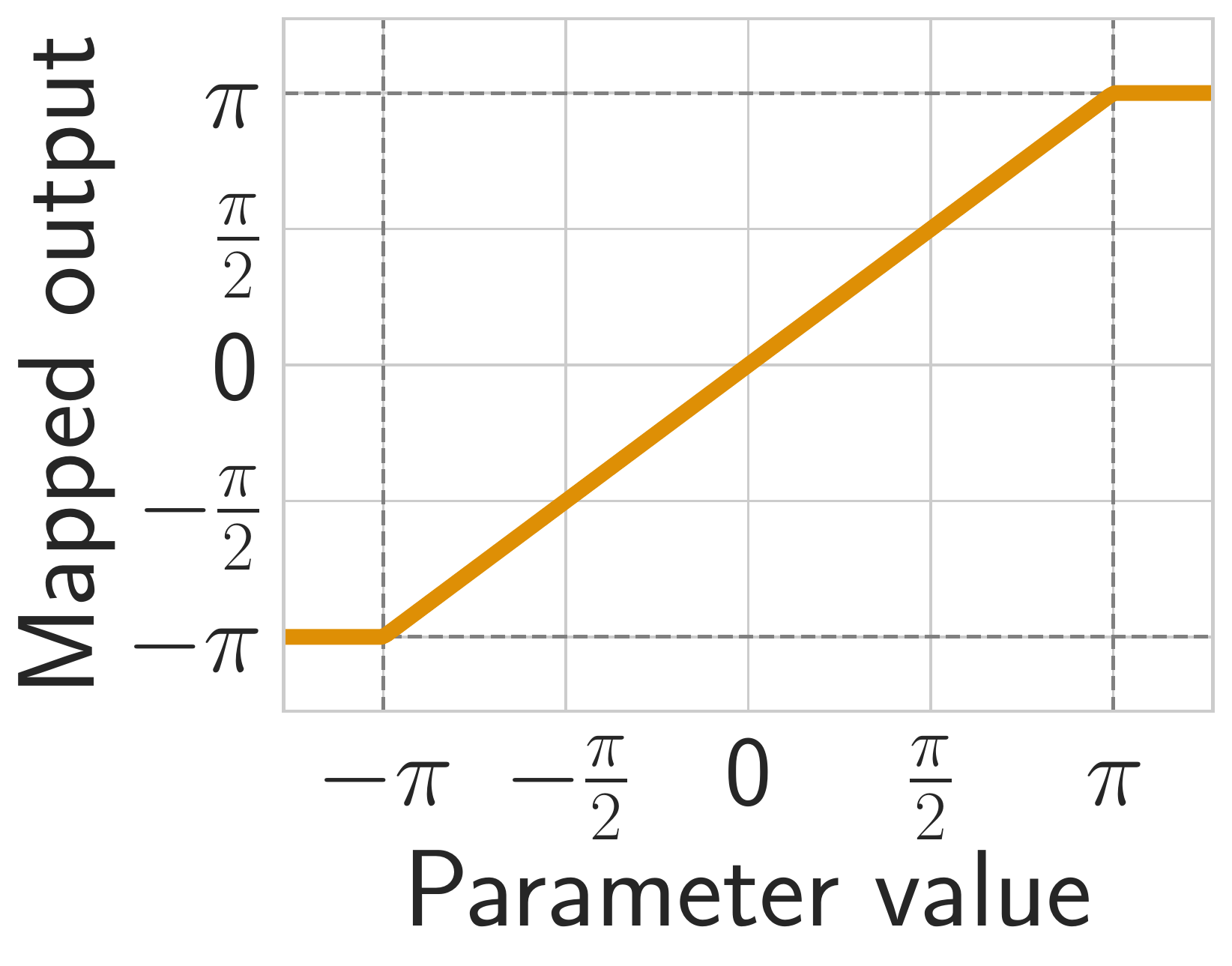}
         \caption{Clamp}
         \label{fig:func-clamp}
     \end{subfigure}
     \begin{subfigure}[t]{0.16\textwidth}
         \centering
         \includegraphics[width=\textwidth]{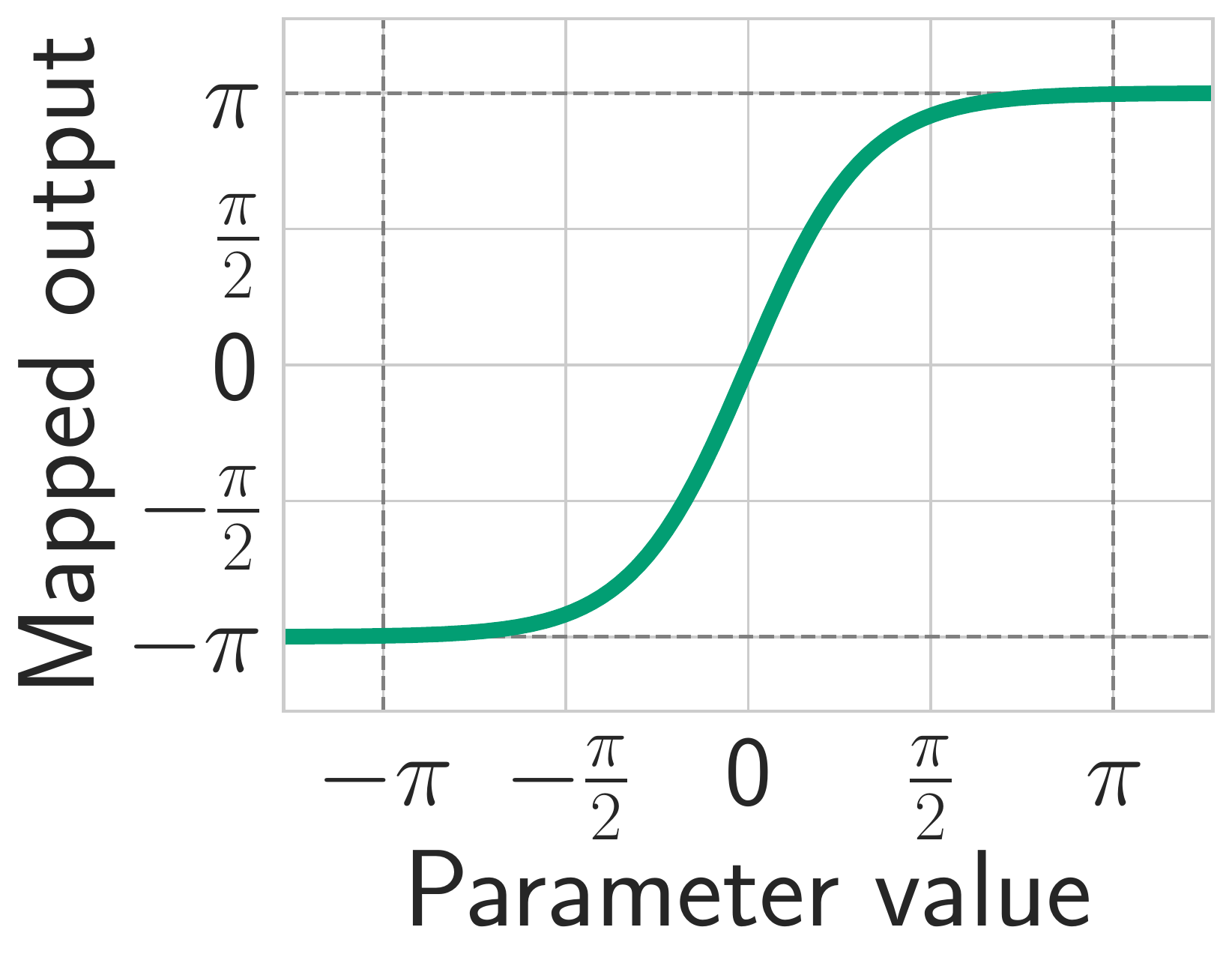}
         \caption{Tanh}
         \label{fig:func-tanh}
     \end{subfigure}
     \begin{subfigure}[t]{0.16\textwidth}
         \centering
         \includegraphics[width=\textwidth]{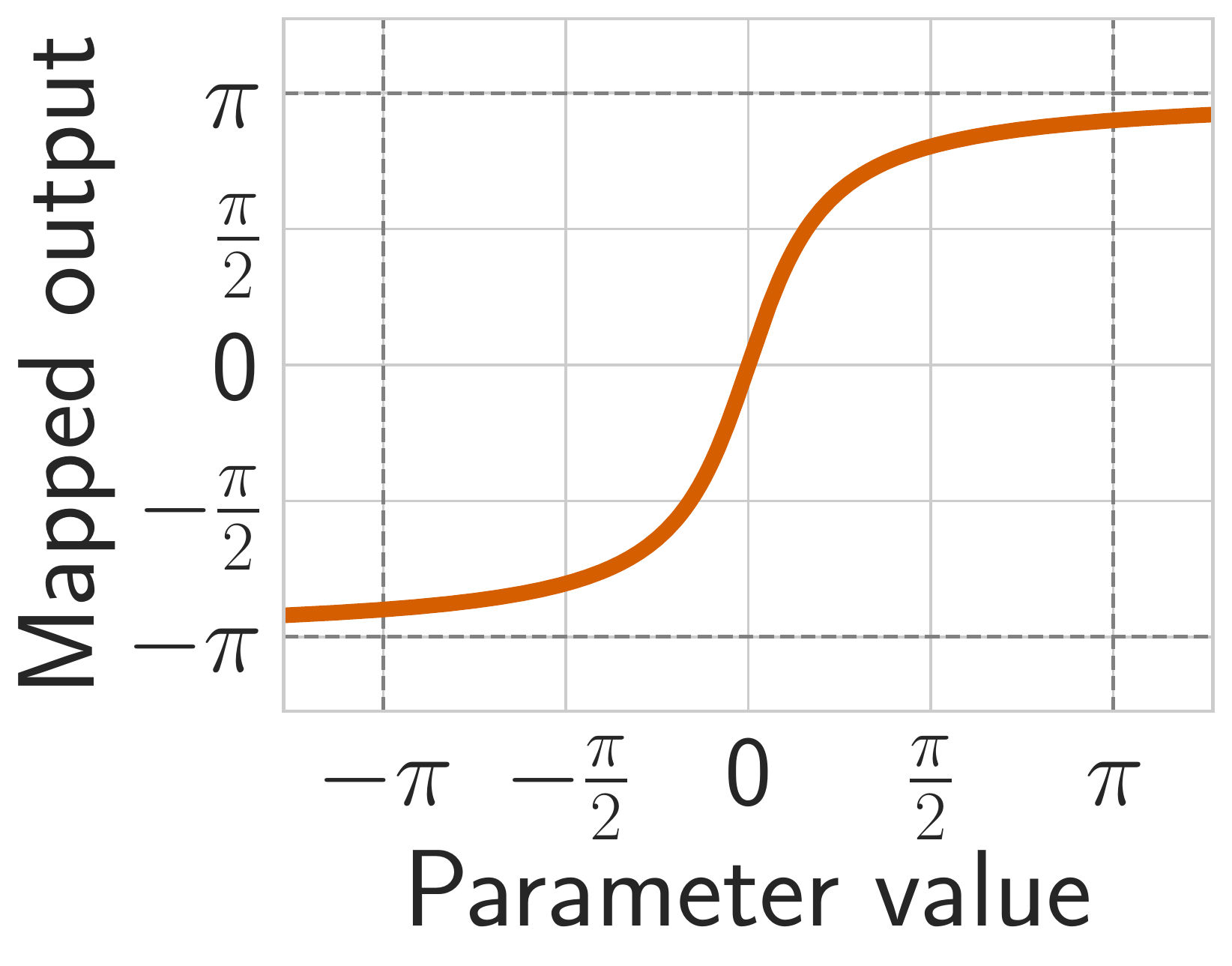}
         \caption{Arctan}
         \label{fig:func-arctan}
     \end{subfigure}
     \begin{subfigure}[t]{0.16\textwidth}
         \centering
         \includegraphics[width=\textwidth]{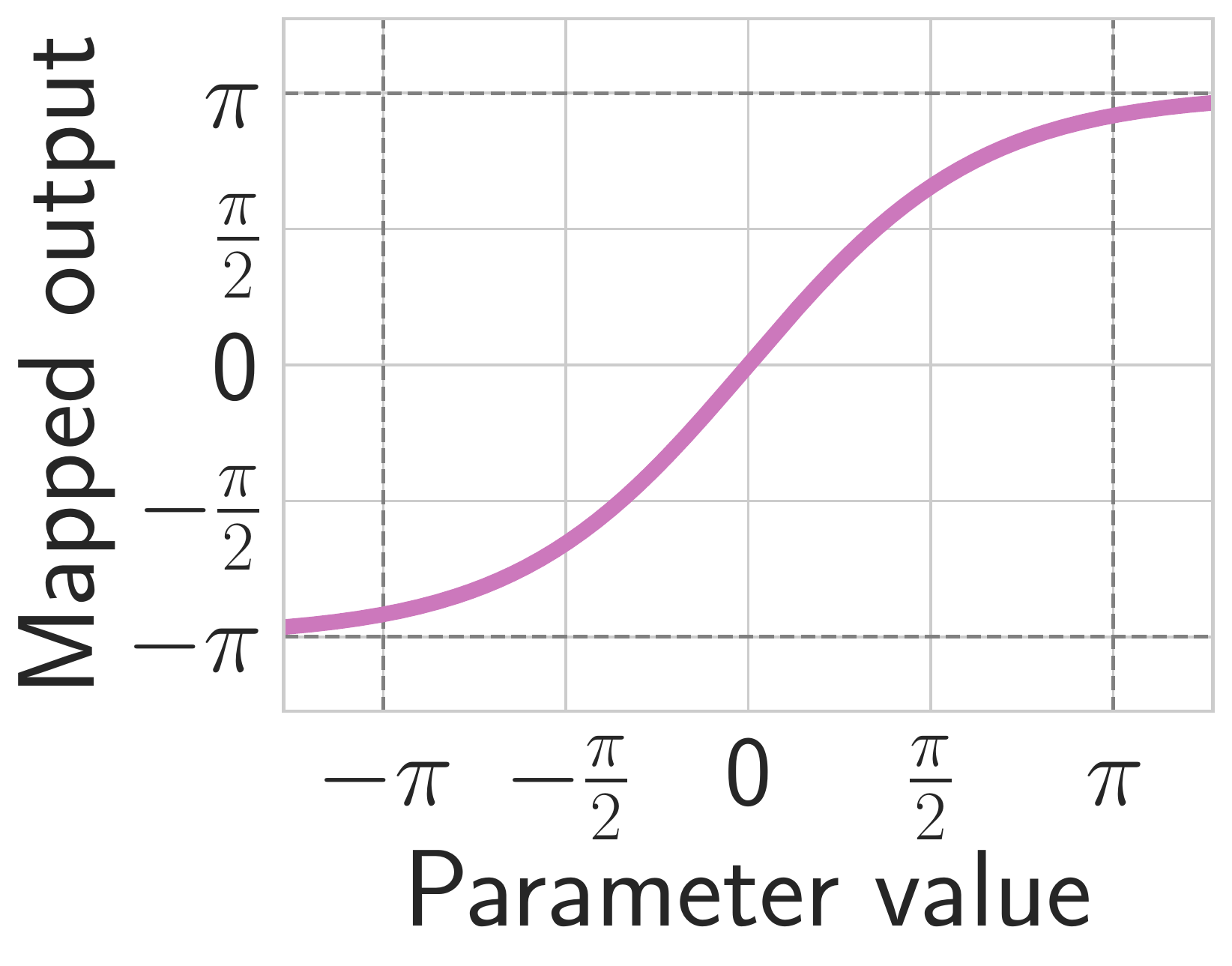}
         \caption{Sigmoid}
         \label{fig:func-sigmoid}
     \end{subfigure}
     \begin{subfigure}[t]{0.16\textwidth}
         \centering
         \includegraphics[width=\textwidth]{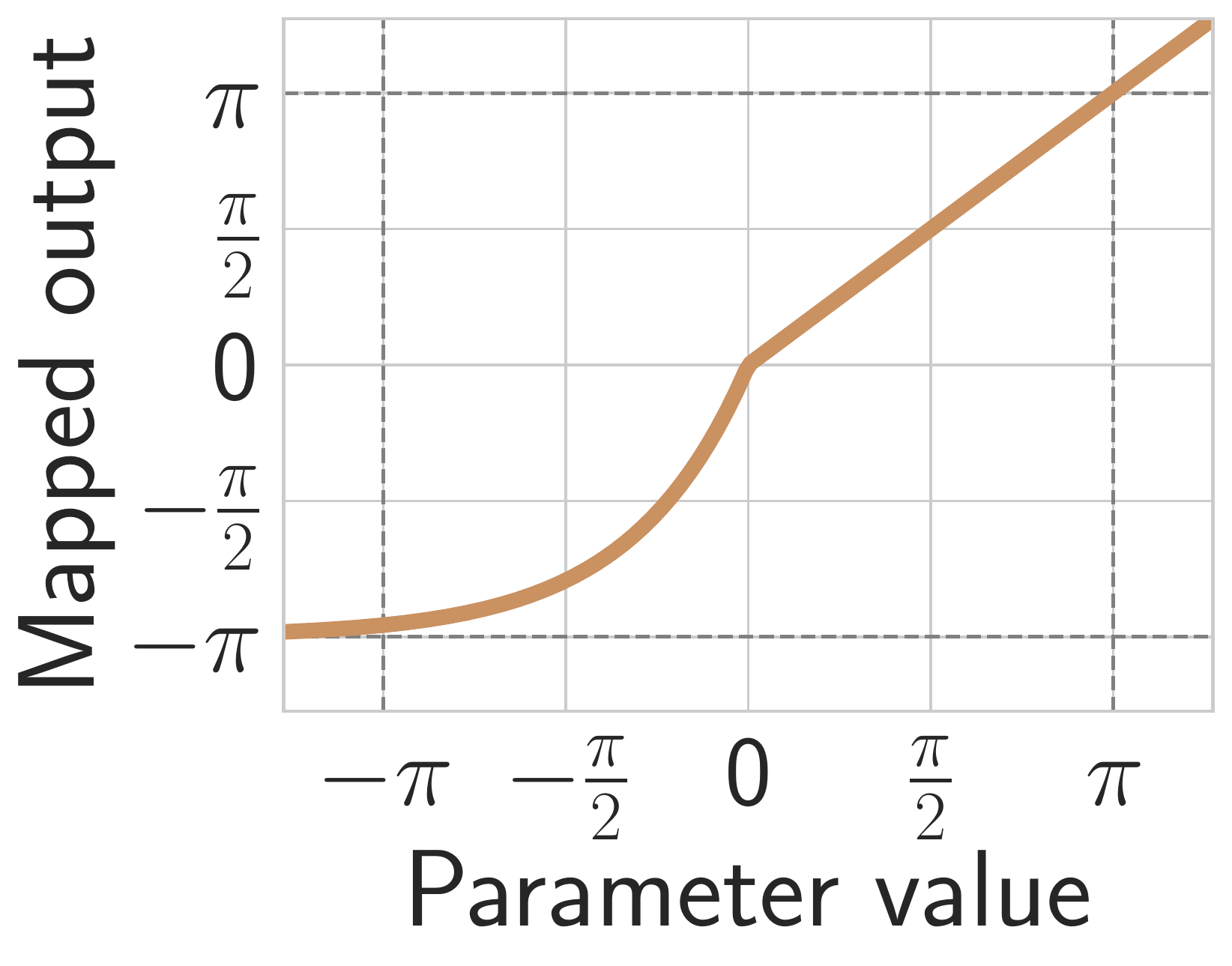}
         \caption{Elu}
         \label{fig:func-elu}
     \end{subfigure}
        \caption{Re-Mapping functions}
        \label{fig:mapping-functions}
\end{figure*}

In a classical neural network the weights that characterize the connections between the neurons are represented by a vector $\vec{\theta} \in \mathbb{R}^n$. Once the neural network has been evaluated and the back-propagation has been performed, we obtain the gradient of the loss function with respect to the parameters $\vec{\theta}$, namely $\nabla_{\vec{\theta}} \mathcal{L}(\vec{\theta})$. The weights are then updated as follows: 

\begin{equation}\label{update_rule}
    \vec{\theta}_i = \vec{\theta}_{i-1} - \alpha \nabla_{\vec{\theta}_{i-1}} \mathcal{L}(\vec{\theta}_{i-1})
\end{equation}
This update step relies on the fact that the space in which we are moving to minimize the loss function is $\mathbb{R}^n$, and thus every value of the real number line can be assumed by the weights.
The same is not true in the case of quantum variational circuits, where, as previously discussed, the parameters $\vec{\theta}$ represent rotations around one of the three axes in the space of the Bloch sphere. 
Since a rotation of angle $\theta$ around a generic direction $\hat{v}$ has a period of $2\pi$, meaning $R(\hat{v}, \theta) = R(\vec{v}, \theta + 2\pi)$, it follows that for a parameter $\theta$ that is close enough to the boundaries of the period, updating the value according to (\ref{update_rule}) may result in making the parameter end up in an adjacent period and thus assume an unexpected value, possibly opposite to the desired direction.
We can conclude that the natural space where the parameters of a quantum variational circuit should be taken is $[\phi, \phi + 2 \pi ]$, with $\phi \in \mathbb{R}$. To center the interval around the value 0 we pick $\phi = -\pi$. This way the parameters of a quantum variational circuit are $\vec{\theta} \in [-\pi, \pi]^n$.
In order to constrain the parameters in the interval $[-\pi, \pi]$ we apply a \textit{mapping function} to the weights. A mapping function is any function of the type 
    
\begin{equation}
    \varphi : \mathbb{R} \to [a, b]
\end{equation}
    
\noindent that maps the real line to a compact interval. In our case we are interested in $ a = -b = - \pi$, so that $\varphi : \mathbb{R} \to [-\pi, \pi]$. Introducing the mapping function means that, whenever a forward pass in performed, every rotation gate depending of a set of angles $\theta = (\theta_x, \theta_y, \theta_z)$ will first have its parameters remapped.

\begin{center}
    \begin{quantikz}[column sep=1cm]
    & \gate{R(\varphi(\theta))} &  \qw
    \end{quantikz}
\end{center}

\noindent We want to emphasize that because the parameters $\vec{\theta}$ will be constrained in the forward pass, they are always free to take values from the real line. This way we can rewrite the basic update rule as follows:
    
\begin{equation}
    \vec{\theta}_i = \vec{\theta}_{i-1} - \alpha \nabla_{\vec{\theta}_{i-1}} \mathcal{L}( \varphi( \vec{\theta}_{i-1} ) )
\end{equation}

\noindent This shows that no mapping function is applied to the weights during the update step, while it is during the forward pass, represented by the loss function.\\
    
\noindent The mapping functions we decided to use in the experiments are some of the most common functions normally used to map the real numbers into a compact interval, properly scaled in such a way that the output interval is $[-\pi, \pi]$. First, we introduce the identity function, which serves as our baseline without a weight constraint. This is equal to applying no mapping at all, as can be seen in Figure \ref{fig:func-wo-constraint}.

\begin{equation} \label{id}
    \varphi_1(\theta) = \theta
\end{equation}

\noindent Starting out with the first mapping function in Figure \ref{fig:func-clamp}, we simply clamp all parameter values above $\pi$ to $\pi$ and all values below $-\pi$ to $-\pi$. Within the interval $[-\pi,\pi]$ the identity function is used. 

\begin{equation} \label{hard_clipping}
    \varphi_2(\theta) = 
        \begin{cases}
            -\pi & \text{if $\theta < -\pi$}\\
            \pi &\text{if $\theta > \pi$}\\
            \theta &\text{otherwise}
        \end{cases}
\end{equation}

\noindent Next, we use the hyperbolic tangent function and apply a scaling of $\pi$. This creates a steeper mapping around $\theta=0$ but has a smoother transition near $-\pi$ and $\pi$, as shown in Figure \ref{fig:func-tanh}
 
\begin{equation} \label{tanh}
    \varphi_3(\theta) = \pi \tanh(\theta)
\end{equation}

\noindent After some initial tests, we wanted to explore the impact of how fast the function approaches the bounds $-\pi$ and $\pi$. Therefore, we scaled the inverse tangent function with factor 2 in both axes. Its graph is shown in Figure \ref{fig:func-arctan}.

\begin{equation} \label{arctan}
    \varphi_4(\theta) = 2 \arctan(2 \theta)
\end{equation}

\noindent Furthermore, we tested a modified sigmoid function, which is not as steep as the previous two and approaches the bounds faster than $\varphi_2$ but slower than $\varphi_3$. To archive the graph in Figure \ref{fig:func-sigmoid}, we needed to scale the function with factor $2\pi$ and shift it by $-\pi$. 
    
\begin{equation} \label{logistic}
    \varphi_5(\theta) = \frac{2\pi}{1+e^{-\theta}} -\pi
\end{equation}

\noindent Lastly, we tested the Exponential Linear Unit or ELU as a asymmetric function, expecting that it should not work as good as the previously mentioned functions. To make sure that the function at least converges to the lower bound $-\pi$ for $\theta < 0$, we set its $\alpha$ parameter to $\pi$. Its graph can be seen in Figure \ref{fig:func-elu}.  

\begin{equation} \label{elu}
    \varphi_6(\theta) = 
        \begin{cases}
            \pi(e^\theta-1) &\text{if $\theta < 0$}\\
            \theta &\text{otherwise}
        \end{cases}
\end{equation}

\subsection{Variational Circuit Architecture}
\label{sec:variational-classifier}
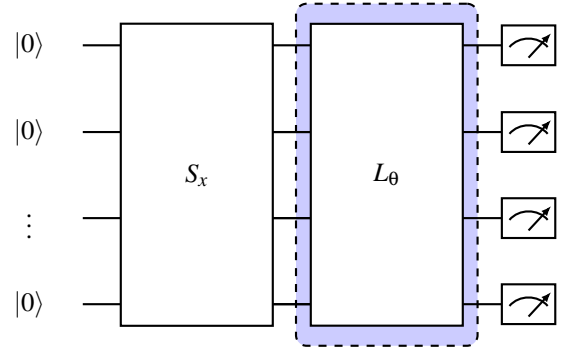
\begin{figure}
    \centering
    \begin{quantikz}
        \ket{0} &  & \gate[wires=4][2cm]{S_x} & \gate[wires=4][2cm]{L_\theta} \gategroup[4,steps=1,style={dashed,
rounded corners,fill=blue!20, inner xsep=2pt},
background]{{\sc }}& \meter{} \\
       \ket{0} & & & \targ{} & \meter{} \\
       \vdots &  & & \qw & \meter{} \\
       \ket{0} & & & \qw & \meter{}
    \end{quantikz}
    
    \caption{Abstract variational quantum circuit used in this work. Dashed blue area indicates repeated layers.}
    \label{fig:abstract_circuit}
\end{figure}

The variational quantum circuit used in the variational classifier consists of 3 parts (Figure \ref{fig:abstract_circuit}). 

\subsubsection*{State preparation}
Initially, we start with the state preparation $S_x$, where the feature vector is embedded into the circuit. Here, real values of the feature vector are mapped from the Euclidean space to the Hilbert space. There are many ways to embed data into a circuit and many different state preparation methods. After a small scale pre-test, we determined that the Angle Embedding yielded the best results on both the Iris and Wine dataset.
\textit{Angle Embedding} is a simple technique to encode $n$ real-valued features into $n$ qubits. Each of the qubits are initialized as $\ket{0}$ and then rotated by some angle around either the $x$- or $y$-axis. The chosen angle corresponds to the feature amplitude to be encoded. Usually single axis rotational gates, are applied, as seen in the example 

\begin{center}
    \begin{quantikz}[column sep=1cm]
    & \gate{R_i(x_j)} &  \qw
    \end{quantikz}
\end{center}

\noindent if we call $x_j \in \mathbb{R}$ the $j$-th feature to be embedded and $R_i, \text{ with } i \in \{X, Y\}$ the rotational gate around the $i$-th axis, then the embedding is performed by simply applying the gate represented below. Note that the $z$-axis is left out: since the initial qubits are in the $\ket{0}$ state, which correspond in the Bloch sphere to vectors along the $z$-axis, rotations around it naturally have no effect and no information would be encoded.

\subsubsection*{Variational Layers}

The second part of the circuit $L_\theta$, hereafter called layers, consists of repeated single qubit rotations and entanglers (in Figure \ref{fig:abstract_circuit} everything contained within the dashed blue area is repeated $L$ times, where $L$ is the layer count). Specifically, we use a layer architecture inspired by the circuit-centric classifier design \cite{Schuld_2020}. All circuits presented in this paper use three single qubit rotation gates and CNOT gates as entanglers. Here, we show an example of the first layer $L_\theta$ of a three qubit variational classifier.\\

\begin{adjustbox}{width=\linewidth}
\begin{quantikz}
     \qw& \gate{RZ(\theta_0^0)} & \gate{RY(\theta_0^1)}& \gate{RZ(\theta_0^2)} & \ctrl{1} & \qw      & \targ{} & \qw   \\
     \qw&\gate{RZ(\theta_1^0)} & \gate{RY(\theta_1^1)}& \gate{RZ(\theta_1^2)} & \targ{}  & \ctrl{1} & \qw    & \qw    \\
     \qw&\gate{RZ(\theta_2^0)} & \gate{RY(\theta_2^1)}& \gate{RZ(\theta_2^2)} & \qw      & \targ{}  & \ctrl{-2}& \qw 
\end{quantikz}
\end{adjustbox}
\\\\

\noindent In the circuit above $\theta_i^j$ denotes a trainable parameter where $i$ represents the qubit index and $j \in \{0,1,2\}$ the index of the single qubit rotation gate. For clarity, we omitted the index $l$ which denotes the current layer in the circuit. In each layer, the target bit of the CNOT gate is given by $(i + l)\ \text{mod}\ n$. For example, in the first layer $l=1$ the control and target qubits are zero qubits apart and are direct neighbors in a circular manner. In the next layer $l=2$, the control and target qubits are one qubit apart. Continuing the example, the index of the target qubit of qubit $i=0$ is $(0 + 2)\ \text{mod}\ 3 = 2$. 

\subsubsection*{Measurement}
The last part of the circuit architecture is the measurement. We measure the expectation value in the computational basis ($z$) of the first $k$ qubits where $k$ is the number of classes to be determined. Then a bias is added to each measured expectation value. The biases are also part of the parameters of the VQC and are updated accordingly. Lastly, the softmax of the measurement vector is computed in order to normalize the probabilities.



\subsection{Datasets}
In this section, we present the datasets that were used to assess the variational classifier from section \ref{sec:variational-classifier}. We chose two popular datasets that pose a simple supervised learning classification task.

\subsubsection{Iris Dataset}
\label{sec:iris}
The Iris dataset originates from an article by Fisher \cite{fisher1936use} in 1936. Since then it has been frequently referenced in many publications to this day. We are using a newer version of the dataset, where two small errors were fixed in comparison to the original article. The dataset includes $3$ classes with $50$ instances each, where each class refers to one of the following types of iris plants: Iris Setosa, Iris Versicolour, Iris Virginica. Each feature vector contains sepal length (in cm), sepal width (in cm), petal length (in cm) and petal width (in cm) respectively. The Iris dataset is a particularly easy dataset to classify, especially when considered that the first two classes are linearly separable. The latter two classes are not linearly separable.

\subsubsection{Wine Dataset}
\label{sec:wine}
The Wine dataset was created in July 1991 as a result of a chemical analysis of wines of three different cultivars grown in the same region in Italy \cite{parvus}. Originally, they investigated $30$ constituents but they were then lost to time and only a smaller version of the dataset still remains. The remaining dataset includes $3$ classes which represent the different cultivars, a total of $178$ instances containing $13$ features each: Alcohol, Malic acid, Ash, Alcalinity of ash, Magnesium, Total phenols, Flavanoids, Nonflavanoid phenols, Proanthocyanins, Color intensity, Hue, OD280/OD315 of diluted wines and Proline. The dataset is not balanced and the instances are distributed among the classes as follows: class 1: 59, class 2: 71, class 3: 48.

\begin{table}[t]
\begin{adjustbox}{width=\linewidth}
\begin{tabular}{|l|ll|}
\hline
Parameter        & Iris Dataset             & Wine Dataset             \\ \hline
Learning Rate    & 0.0201                   & 0.0300                   \\
Weight Decay     & 0.0372                   & 0.0007                   \\
Batch Size       & 9                        & 18                       \\
Embedding        & Angle (X-Axis) & Angle (Y-Axis) \\
Number of Layers & 8                        & 9                        \\ \hline
\end{tabular}
\end{adjustbox}
\caption{Hyperparameters used for each dataset}
\label{tab:hyperparameters}
\end{table}
\begin{table*}[t]
    \centering
    \begin{adjustbox}{width=\linewidth}
    \begin{tabular}{|r|cccccc|} 
     \hline
     Dataset        & w/o Re-Mapping & Clamp & Tanh  & Arctan    & Sigmoid   & ELU \\ \hline
     Iris           & $0.953\pm 0.024$                  & $0.953\pm 0.024$  & $0.953\pm 0.024$  & \pmb{$0.957\pm 0.023$}& $0.950\pm 0.025$      & $0.943\pm 0.026$\\ 
     Wine           & $0.617\pm 0.071$                  & $0.622\pm 0.071$  & $0.706\pm 0.067$  & $0.667\pm 0.069$      & \pmb{$0.717\pm 0.066$}& $0.694\pm 0.067$\\ \hline
    \end{tabular}
    \end{adjustbox}
    \caption{Test Accuracy of tested mapping functions with 95\% confidence interval}
    \label{tab:test-acc}
\end{table*}
\subsection{Training}
\label{sec:training}
We started the training of the variational classifiers with a hyperparameter search using Bayesian optimization. We trained approximately 300 runs per dataset, optimizing the test accuracy. All runs in the hyperparamter search were executed without weight constraints. The hyperparameters of the best run for each dataset can be found in Table \ref{tab:hyperparameters}. We then ran 20 runs for every mapping function on the Iris dataset and 10 runs each on Wine dataset, each with the before mentioned hyperparameters.
The initialization of the weights is important when using weight constraints, since the functions defined in Section \ref{sec:weight-constraints} are centered around $0$. Therefore, we initialized the weights close to $0$ in the range of [$-0.01$, $0.01$].
For the training we used Cross Entropy Loss and Adam Optimizer.

\begin{table}[t]
    \begin{adjustbox}{width=\linewidth}
    \begin{tabular}{|ccccccc|}
    \hline
    \multicolumn{7}{|c|}{Iris Dataset}                                                                                        \\ \hline
    \multicolumn{1}{|c|}{Samples} & w/o Re-Mapping & Clamp         & Tanh           & Arctan        & Sigmoid & ELU   \\ \hline
    \multicolumn{1}{|c|}{120}               & 0.693          & 0.700         & 0.903          & \pmb{$0.913$} & 0.837   & 0.840 \\
    \multicolumn{1}{|c|}{240}              & 0.883          & 0.887         & \pmb{$0.937$}  & 0.927         & 0.927   & 0.910 \\
    \multicolumn{1}{|c|}{480}              & 0.933          & \pmb{$0.937$} & 0.927          & 0.933         & 0.930   & 0.920 \\ \hline
    \multicolumn{7}{|c|}{Wine Dataset}                                                                                        \\ \hline
    \multicolumn{1}{|c|}{Samples} & w/o Re-Mapping & Clamp         & Tanh           & Arctan        & Sigmoid & ELU   \\ \hline
    \multicolumn{1}{|c|}{568}              & 0.372          & 0.372         & \pmb{$0.533$} & 0.522         & 0.489   & 0.511 \\
    \multicolumn{1}{|c|}{994}             & 0.511          & 0.517         & \pmb{$0.639$} & 0.539         & 0.583   & 0.533 \\
    \multicolumn{1}{|c|}{1846}             & 0.572          & 0.572         & \pmb{$0.656$}  & 0.639         & 0.628   & 0.628 \\ \hline
    \end{tabular}
    \end{adjustbox}
    \caption{Accuracy (Validation) of tested mapping functions during convergence}
    \label{tab:convergence}
\end{table}

\section{\uppercase{Experiments}}
\begin{figure*}[ht]
    \centering
    \includegraphics[width=\linewidth]{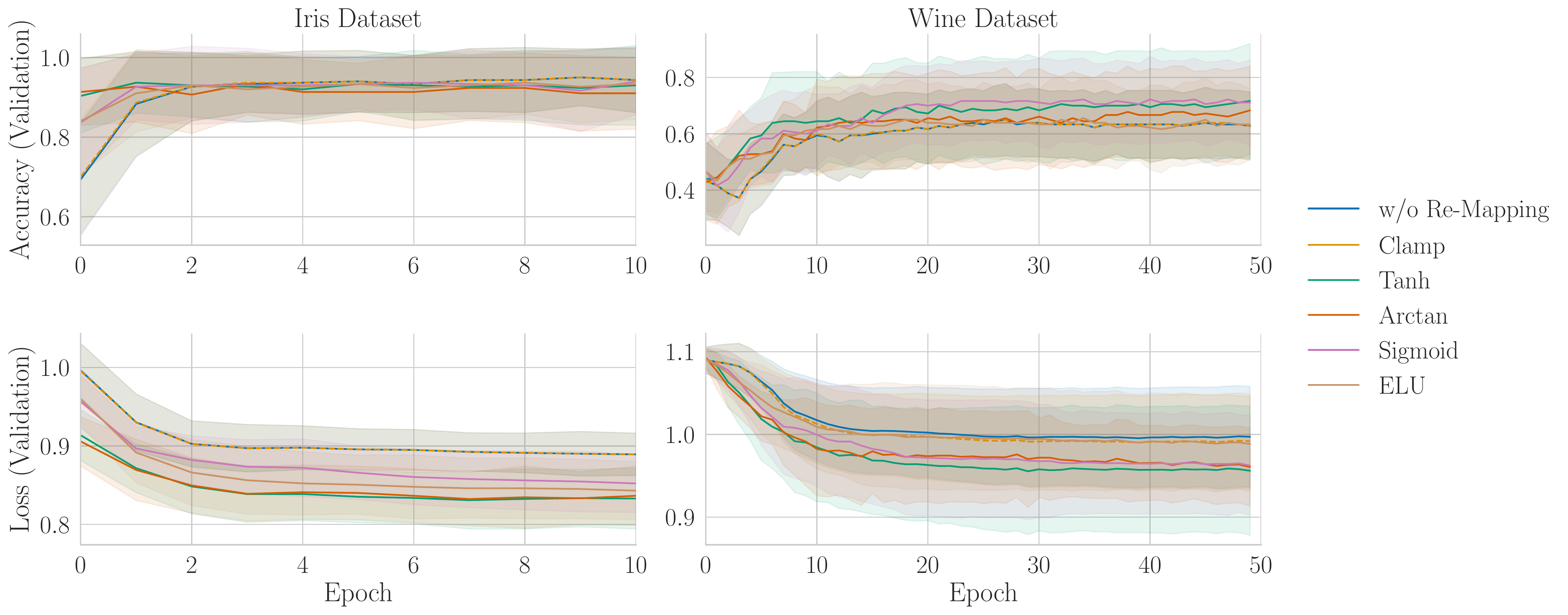}
    \caption{Validation curves for datasets Iris and Wine. In each epoch the algorithm processes $80\%$ of the total samples of each dataset for training. This accounts to $120$ samples for the Iris dataset and $142$ samples for the Wine dataset. }
    \label{fig:results_valid}
\end{figure*}

In this section we present our conducted experiments on the Iris and Wine datasets. To show the impact of the re-mapping functions from Section \ref{sec:weight-constraints}, we chose simple supervised learning tasks using a quantum variational classifier. However, in principle our approach can be used for any scenario that relies on a variational quantum circuit, for example Quantum Reinforcement Learning, QAOA, and more. We trained the classifier on two different datasets Iris (Section \ref{sec:iris}) and Wine (Section \ref{sec:wine}). All training related details can be found in Section \ref{sec:training}. We first investigated the convergence behavior with and without the weight re-mapping. This is an important topic, especially for quantum machine learning, since the use of quantum hardware is much more expensive than classical hardware at the time of writing. In Figure \ref{fig:results_valid} and Table \ref{tab:convergence} we show that faster convergence can be achieved by using weight re-mapping in all tested settings. With a faster convergence we can potentially save on compute which results in less power consumption and $\text{C0}_2$ emissions. We also investigated overall test accuracy in Table \ref{tab:test-acc}, where we can see minor improvements for the Iris dataset and a $10\%$ improvement for the Wine dataset, compared to using no weight re-mapping.

\subsection{Iris Dataset}
In the following, we present the results of our experiments on the Iris dataset. In a pre-test we have found that Angle Embedding with rotations around the $x$-axis works best for the classification of the Iris dataset, but also requires more qubits than other embedding methods like Amplitude Embedding \cite{mottonen}. It should be noted, that Angle Embedding with rotations around the $y$-axis should, in theory, perform similarly. Angle Embedding with rotations around the $z$-axis does not work at all, since the rotations are invariant to the expectation value of the measurement w.r.t. $z$-axis. We then ran a quick hyperparameter search on the architecture without a re-mapping function, determining the hyperparameters listed in Table \ref{tab:hyperparameters}. We wanted to see if we can further improve the convergence of an architecture with an already good set of hyperparameters, just by introducing weight re-mapping. We trained the model with and without weight re-mapping with $20$ different seeds on the dataset. The whole training curves can be found in Figure \ref{fig:results_valid} and the convergence behavior can be seen in detail in Table \ref{tab:convergence}, where we picked three points during convergence and compared the accuracies. We observe a increase of over $20\%$ using the Arctan re-mapping compared to using no re-mapping after $120$ samples. Similarly, after $240$ samples the model Tanh mapping has already converged and resulted in over $5\%$ higher accuracy. As the model without re-mapping reaches convergence after $480$ samples, we see practically no difference between the approaches. Theses differences can be attributable to the re-mapping functions. In this particular case the optimal weights are close to zero and especially Tanh and Arctan are very steep in this area. One possibility is that the re-mapping kept the values closer to this area resulting in finding the weights faster. We also investigated if the re-mapping process has any impact on the overall test accuracy, where the results can be found in Table \ref{tab:test-acc}. As expected, the test accuracy did not decrease when using weight re-mapping for the Iris dataset. It even reached a minimally higher test accuracy which is due to more time for fine-tuning after the earlier convergence. In this setting, Arctan is our overall preferred choice because of its fast convergence and second best test accuracy.

\subsection{Wine Dataset}
The results of our experiments on the Wine dataset are presented below. In a preliminary test, we discovered that, in contrast to the Iris dataset, Angle Embedding with rotations around the $y$-axis works best. In theory, Angle Embedding with rotations around the $x$-axis should perform similarly, since we are measuring the expectation value w.r.t. $z$-axis and rotations around the $z$-axis do not work at all, since the rotations are invariant to the expectation value of the measurement. Like the architecture used for the Iris dataset, Angle Embedding outperformed more efficient embedding methods like Amplitude Embedding \cite{mottonen}. The architecture without a re-mapping function was then also subjected to a small scale hyperparameter search, yielding the hyperparameters listed in Table \ref{tab:hyperparameters}. We were interested in determining whether adding weight re-mapping may further enhance the convergence of an architecture with a good set of hyperparameters. On the dataset, we trained the model both with and without weight re-mapping using $20$ seeds. The entire training curves are shown in Figure \ref{fig:results_valid}, and Table \ref{tab:convergence} shows the convergence behavior in detail, comparing the accuracies at three points during convergence. From the results, we can see that Tanh is outperforming all other tested settings during convergence and models with weights re-mapping generally converging faster than the baseline. At $568$ samples Tanh has a $15\%$ higher accuracy compared to the model without weight re-mapping. After $994$ samples Tanh is almost fully converged, while the baseline only converges long after $1846$ samples. We believe that the cause of this effect is the same as that mentioned in the section above. These differences are brought on by the re-mapping operation. In this case, the ideal weights are almost zero, and Tanh is very steep in this region, mapping the values close to zero and enabling a more detailed search in this region, ultimately resulting in a faster convergence. Additionally, Table \ref{tab:test-acc} shows that the weight re-mapping not only leads to a faster convergence but also, in this case, a $10\%$ improvement in test accuracy, closely followed by Tanh. Tanh is our first preference in this situation because of its quick convergence and second-best test accuracy.

\section{\uppercase{Conclusion}}
\label{sec:conclusion}
In this work, we introduced weight re-mapping for variational quantum circuits, a novel method for accelerating convergence using quantum variational classifiers as an example. In order to test our classifier on two datasets — Iris and Wine — we first conducted an initial hyperparameter search to identify a good set of parameters without re-mapping. After that, we introduced our weight re-mapping approach for evaluation.
Our experiments show that weight re-mapping can improve convergence in all tested settings. Faster convergence may allow us to lower the amount of compute we need, which will reduce power consumption and $\text{C0}_2$ emissions. This is crucial, especially for quantum machine learning, where access to quantum hardware is quite expensive.
Furthermore, for the Wine dataset, we were able to demonstrate that using weight re-mapping improved test accuracy by $10\%$ compared to using unmodified weights. For both settings, we would recommend a re-mapping function that is steep around the initialization point, such as Arctan and Tanh. 
There are additional opportunities based on our findings. This includes testing other datasets, with more features or more complex classification problems. We also see opportunities to develop specialized optimizers that take into account the unique characteristics of parameters in quantum circuits. Finally, more research is required to determine the impact of weight re-mapping on other tasks and fields, such as Quantum Reinforcement Learning.

\vfill

\bibliographystyle{apalike}
{\small
\bibliography{main}}

\end{document}